\documentclass{ws-mpla}

\usepackage{amssymb}
\usepackage{amsmath}
\usepackage{amsfonts}
\usepackage{graphicx}
\begin{document}

\markboth{C. A. Dominguez, L. A. Hernandez, K. Schilcher, H. Spiesberger}
{Tests of quark-hadron duality in tau-decays}

\catchline{}{}{}{}{}

\title{
TESTS OF QUARK-HADRON DUALITY IN TAU-DECAYS
}

\author{C. A. DOMINGUEZ} 
\address{Centre for Theoretical and Mathematical Physics, 
and Department of Physics, University of Cape Town, 
Rondebosch 7700, South Africa, and 
\\
National Insitute for Theoretical Physics, 
Private Bag X1, Matieland 07602, South Africa} 

\author{L. A. HERNANDEZ}
\address{Centre for Theoretical and Mathematical Physics, 
and Department of Physics, University of Cape Town, 
Rondebosch 7700, South Africa}

\author{K. SCHILCHER\footnote{Speaker}}
\address{PRISMA Cluster of Excellence, 
Institut f\"{u}r Physik,
Johannes Gutenberg-Universit\"{a}t, 
D-55099 Mainz, Germany, and 
\\
Centre for Theoretical and Mathematical Physics, 
and Department of Physics, University of Cape Town, 
Rondebosch 7700, South Africa, and 
\\ 
National Insitute for Theoretical Physics, 
Private Bag X1, Matieland 07602, South Africa} 

\author{H. SPIESBERGER}
\address{PRISMA Cluster of Excellence, 
Institut f\"{u}r Physik,
Johannes Gutenberg-Universit\"{a}t, 
D-55099 Mainz, Germany, and 
\\
Centre for Theoretical and Mathematical Physics, 
and Department of Physics, University of Cape Town, 
Rondebosch 7700, South Africa}


\maketitle

\pub{Received (Day Month Year)}{Revised (Day Month Year)}

\begin{abstract}
\noindent
An exhaustive number of QCD finite energy sum rules for 
$\tau$-decay together with the latest updated ALEPH data 
is used to test the assumption of global duality. Typical 
checks are the absence of the dimension $d=2$ condensate, 
the equality of the gluon condensate extracted from vector 
or axial vector spectral functions, the Weinberg sum rules, 
the chiral condensates of dimensions $d=6$ and $d=8$, as well 
as the extraction of some low-energy parameters of chiral 
perturbation theory. Suitable pinched linear integration 
kernels are introduced in the sum rules in order to suppress
potential quark-hadron duality violations and experimental 
errors. We find no compelling indications of duality violations 
in hadronic $\tau$-decay in the kinematic region above 
$s\simeq2.2$ GeV$^{2}$ for these kernels.
\keywords{QCD; Sum Rules; Tau Decays.}
\end{abstract}

\ccode{PACS Nos.: 12.38.Lg, 11.55.Hx, 13.35.Dx}


\section{Introduction}

Tau decay constitutes the ideal laboratory to test subtle effects 
of QCD in the intermediate energy region which is still accessible 
to the perturbation series and the operator product expansion 
(OPE). To present the case, we collect here the results that may 
be obtained with minimal assumptions. The only information we use 
is the one that has been calculated explicitly, i.e.\ the known 
coefficients of a perturbative series and the Wilson coefficients 
of the OPE. The hypothesis that this approach works is known to 
be violated in practice, for instance, by the factorial growth 
of the coefficients and the presence of duality violations (DV). 
We assume that the asymptotic behavior sets in at fairly high
perturbative orders so that it does not affect phenomenological 
applications considered here. As for possible DV, we minimize 
their impact by restricting our analysis to pinched finite energy 
sum rules (FESR) involving spectral function moments of low mass 
dimensions. Pinching serves two purposes, it reduces the 
contribution near the positive real axis in the contour integral
of the QCD correlator and it reduces the experimental errors of 
the $\tau$ spectral functions. Our approach is orthogonal to a 
number of recent papers\cite{Boito1,Pich1} where an ansatz for 
DV is made based on large-$N_{c}$ QCD  and Regge models with the 
aim to identify and quantify DV. For this purpose, higher 
dimensional spectral moments must be considered. We want to
demonstrate that for the low dimensional moments and duality 
radii larger than about 2.2 GeV$^{2}$, there is little or no 
evidence for the existence of DV given the experimental errors 
of the ALEPH data\cite{ALEPH}. To reach such a conclusion it is 
mandatory to consider all sum rules where the answer is known
beforehand. These include the Weinberg and related sum rules for 
the chiral correlator and sum rules for the separate vector ($V$) 
and axial vector ($A$) correlator. For example, by incorporating 
the fact that there is no dimension two operator in QCD we can 
set up simple pinched FESR to demonstrate that the gluon 
condensate is equal and positive for the $V$ and $A$ correlators. 
The following analysis is a summary of two of our recent 
papers\cite{DHSS1,DHSS2}.


\section{QCD finite energy sum rules}

The strangeness conserving hadronic spectral functions in 
$\tau$-decay are related to the (charged) $V$ and $A$ current 
correlators
\begin{align}
\Pi_{\mu\nu}^{VV}(q^{2}) 
&= 
i\int d^{4}x\;e^{iqx}
\langle0|T(V_{\mu}(x)V_{\nu}^{\dagger}(0))|0\rangle 
\label{M1} 
\\
&= 
(-g_{\mu\nu}\;q^{2}+q_{\mu}q_{\nu})\;\Pi_{V}(q^{2}) 
\;, 
\nonumber
\end{align}
\begin{align}
\Pi_{\mu\nu}^{AA}(q^{2}) 
&= 
i\int d^{4}x\;e^{iqx} 
\langle0|T(A_{\mu}(x)A_{\nu}^{\dagger}(0))|0\rangle 
\label{M2} 
\\
&= \; 
(-g_{\mu\nu}q^{2}+q_{\mu}q_{\nu})\;\Pi_{A}(q^{2}) 
- q_{\mu}q_{\nu} \;\Pi_{0}(q^{2}) 
\;, 
\nonumber
\end{align}
where $V_{\mu}(x) = \bar{u}(x)\gamma_{\mu}d(x)$, $A_{\mu}(x) = 
\bar{u}(x)\gamma_{\mu}\gamma_{5}d(x)$ with $u(x)$ and $d(x)$ 
the quark fields. Our starting point is perturbative QCD and the 
operator product expansion (OPE). We write 
\begin{equation}
\Pi^{QCD}(s)=\Pi^{PERT}(s)+\Pi^{OPE}(s) 
\label{M21}
\end{equation}
where $\Pi^{PERT}(s)$ is the perturbation series of massless QCD 
known up to 5 loops\cite{Baikov} and 
\begin{equation}
4\pi^{2}\Pi^{OPE}(Q^{2}) 
\equiv 
\sum_{N=2}^{\infty}\frac{1}{Q^{2N}}\;C_{2N}(Q^{2},\mu^{2})\; 
\langle0|\mathcal{O}_{2N}(\mu^{2})|0\rangle 
\;,
\label{M3}
\end{equation}
and $Q^{2}\equiv-q^{2}$, and the sum is over scalar gauge 
invariant operators of dimension $2, 4, 6, \ldots$. The parameter 
$\mu$ is a renormalization scale separating long from short 
distance physics. Short distances are absorbed in the Wilson 
coefficients $C_{2N}(Q^{2},\mu^{2})$ and long distances in the 
vacuum condensates $\langle O_{2N}(\mu^{2})\rangle$. Quark mass 
effects are completely negligible for the the $u-d$ quark sector 
considered here. The relevant scale of $\tau$-decay is $m_{\tau}$ 
which is considered to be much larger than the scale 
$\Lambda_{QCD}$ associated with non-perturbative effects
beyond the OPE.

Our normalization of the correlators is such that
\begin{equation}
\frac{1}{\pi}\operatorname{Im}\Pi^{PERT}(s) 
= 
\frac{1}{4\pi^{2}}
\left(1+\frac{\alpha_{s}(\mu^{2})}{\pi} + \ldots\right)
\, .
\nonumber
\end{equation}
Note that the term with $N=1$ is absent from the OPE as, in 
massless QCD, there is no gauge invariant operator of dimension 
$d=2$. The question of the absence of such a condensate will, 
however, be checked in our analysis. At dimension $d=4$ the 
contribution from the gluon condensate should be equal for
$V$ and $A$ correlators.
\begin{figure}[h]
\begin{center}
\includegraphics[width=0.50\textwidth]{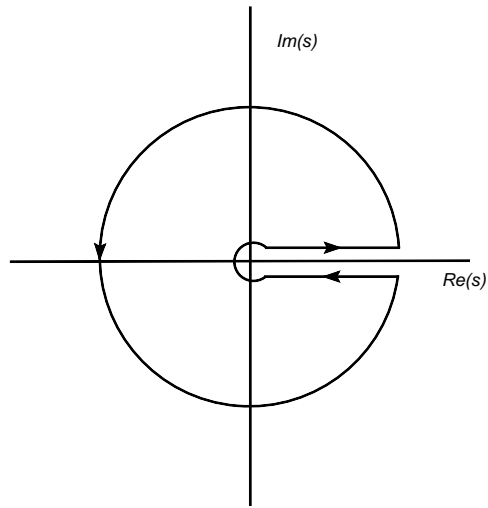}
\caption{\small Integration contour of the FESR.
\label{fig:KS1}}
\end{center}
\end{figure}
The correlators are analytic in the complex $s\ (=q^{2})$ plane 
with a right-hand cut starting at the relevant thesholds. 
Integrating $\Pi(s)$ over the contour of Fig.\ \ref{fig:KS1} 
and applying Cauchy's theorem one obtains the following finite 
energy sum rule (FESR) 
\begin{equation}
\int_{0}^{s_{0}}dsP(s)\rho_{V,A}\left(  s\right) 
= 
-\frac{1}{2\pi i} \oint_{|s|=s_{0}}ds\,P(s)\,\Pi_{V,A}^{QCD}(s) 
+ \text{\emph{Res}}[P(s)\Pi_{V,A}(s)]\text{ ,} 
\label{M4.2} 
\end{equation}
where $\rho_{V,A}\left(s\right)$ are the spectral functions 
measured in $\tau$-decay and $P(s)$ is taken here to be a 
power series 
\begin{equation}
P(s) =
\sum
a_{n}s^{n} \, , 
\quad \quad 
n = \ldots, -2, -1, 0, 1, 2, \ldots
\nonumber
\end{equation}
In the convention of ALEPH\cite{ALEPH}
\begin{equation}
\rho\left(s\right) 
= 
\frac{1}{2\pi^{2}}[v(s),a(s)]_{\text{ALEPH}}\text{ .}
\label{M5}
\end{equation}
To the axial spectral function the pion pole has to be added 
$\rho_{A}^{\text{pion}}(s)=2f_{\pi}^{2}\delta(s-m_{\pi}^{2})$.

We absorb the Wilson coefficients (ignoring radiative corrections) 
into the operators and write the OPE as 
\begin{equation}
\Pi^{\text{OPE}}(Q^{2}) 
= 4\pi^{2}\sum_{N=0}^{\infty}\frac{1}{Q^{2N}}{\mathcal{O}}_{2N} 
\;. 
\label{M42} 
\end{equation}
For $P(s)$ a polynomial, the FESR Eq.\ (\ref{M4.2}) can be written 
as 
\begin{equation}
(-)^{N+1}\,{\mathcal{O}}_{2N+2}^{V,A} 
= 
4\pi^{2}\int_{0}^{s_{0}}ds\,s^{N}
\left[
\begin{array}
[c]{c} 
\rho_{V}(s)\\
\rho_{A}(s)
\end{array}
\right]  +\left[
\begin{array}
[c]{c} 
0 
\\
1
\end{array}
\right] 
2f_{\pi}^{2}-s_{0}^{N+1}M_{2N+2}(s_{0}) 
\,, 
\label{M4} 
\end{equation}
with the PQCD moments $M_{2N+2}(s_{0})$ defined as 
\begin{equation}
M_{N}(s_{0}) 
\equiv 
\int\limits_{0}^{s_{0}}\frac{ds}{s_{0}}
\left[\frac{s}{s_{0}}\right]^{N}4\pi^{2}\frac{1}{\pi}
\operatorname{Im}\Pi^{\text{PQCD}}(s) 
= 
- \frac{1}{2\pi i}\oint\limits_{|s|=s_{0}}\frac{ds}{s_{0}}
\left[\frac{s}{s_{0}}\right]^{N}4\pi^{2} 
\Pi^{\text{PQCD}}(s) 
\nonumber
\end{equation}
for both $V$ and $A$. These moments are dimensionless and 
normalized according to 
\begin{equation}
M_{N} 
= 
\frac{1}{N+1} 
\text{ \ \ \ for \ \ \ } 
\alpha_{s}=0 
\text{\ .} 
\label{M3a} 
\end{equation}
As was proposed some time ago\cite{PINCH}, it is advantageous 
to consider pinched sum rules where the weight vanishes at 
$s_{0}$, the end of the integration range, 
\[
(-)^{N}{\mathcal{O}}_{2N+2}^{V,A} 
= -4\pi^{2}s_{0}^{N}\int_{0}^{s_{0}}ds 
\left[1-\left(\frac{s}{s_{0}}\right)^{N}\right] 
\rho_{V,A}(s)-s_{0}^{N+1} 
\left[M_{0}(s_{0})-M_{2N+2}(s_{0})\right] 
\, .
\]
These sum rules offer two advantages. First they reduce 
significantly the effect of experimental errors which increase 
with $s_{0}$. Second they reduce the contribution of the contour 
integration region near the cut where DV would be most significant.


\section{Chiral sum rules}

The simplest object to study is the chiral correlator 
$\Pi_{V-A}(s)\equiv \Pi_{V}(s)-\Pi_{A}(s)$ because it vanishes 
identically in the chiral limit ($m_{q}=0$), to all orders in 
PQCD. OPE contributions start with the dimension six quark 
condensate,
\begin{equation}
\Pi(Q^{2})|_{V-A}^{\text{OPE}} 
\;=\; 
- \frac{32\pi}{9}\; 
\frac{\alpha_{s}\langle \bar {q}q \rangle^{2}}{Q^{6}} 
\; 
\left\{1+\frac{\alpha_{s}(Q^{2})}{4\pi}\; 
\left[\frac{247}{12}+\mathrm{ln} 
\left(\frac{\mu^{2}}{Q^{2}}\right)  \right]
\right\} 
\; + \mathcal{O}(1/Q^{8}) 
\;, 
\label{M8} 
\end{equation}
The $\alpha_{s}$ corrections were calculated\cite{4qC} in the 
anti-commuting $\gamma_{5}$ scheme and assuming vacuum saturation 
of the four-quark condensate.


\subsection{Weinberg-type sum rules}

\begin{figure}[b]
\begin{center}
\includegraphics[width=0.6\textwidth]{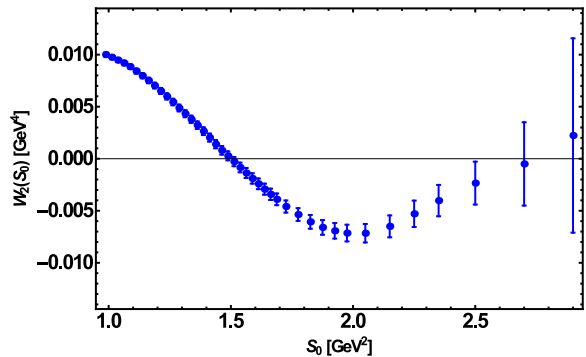}
\caption{\small 
The second WSR, the spectral integral of Eq.\ (\ref{M9}) 
should be zero. 
\label{fig:KS2}}
\end{center}
\end{figure}

We begin the analysis with the second Weinberg sum rule 
(WSR)\cite{WSR}, 
\begin{equation}
\int_{0}^{s_{0}}\;ds\;s\left[\rho_{V}(s)-\rho_{A}(s)\right] 
= 0 
\;,
\label{M9} 
\end{equation}
Using recent ALEPH data\cite{ALEPH} the result is presented in 
Fig.\ \ref{fig:KS2}. 
\begin{figure}[t]
\begin{center}
\includegraphics[width=0.6\textwidth]{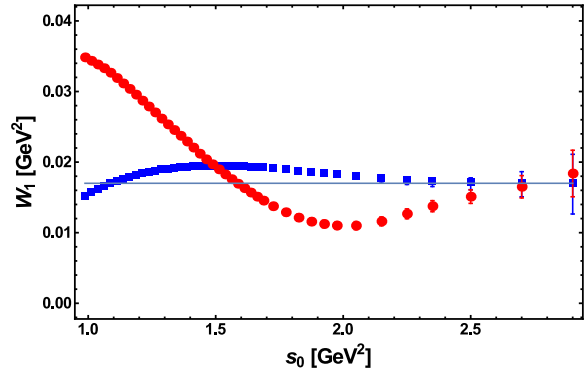}
\caption{\small 
First WSR (dots), Eq.\ (\ref{W26}) and the pinched WSR sum 
rule (squares), Eq.\ (\ref{W27}). The straight line is 
$2f_\pi^2$.  
\label{fig:KS3}}
\end{center}
\end{figure}
The sum rule is not saturated, except possibly near the endpoint 
of the spectrum. Due to the large experimental errors, it is not 
clear to what extent saturation occurs there. The same holds for 
the first Weinberg sum rule. 
\begin{equation}
W_{1}(s_{0})\text{ }\text{: \ \ } 
\int_{s_{thr}}^{s_{0}}ds 
\left[\rho_{V}(s)-\rho_{A}(s)\right] = 2f_{\pi}^{2} 
\;. 
\label{W26}
\end{equation}
As mentioned above, the saturation of the various sum rules can 
be considerably improved by introducing an integration kernel 
that vanishes at the upper limit of integration ($s=s_{0}$). 
Combining the first and second Weinberg sum rules we obtain 
the \emph{pinched first Weinberg sum rule}, 
\begin{equation}
W_{1}(s_{0})\text{ }\text{: \ \ } 
\int_{s_{thr}}^{s_{0}}\;ds\; 
\left(1-\frac{s}{s_{0}}\right)
\left[\rho_{V}(s)-\rho_{A}(s)\right] 
= 2f_{\pi}^{2} 
\;. 
\label{W27}
\end{equation}
It can be seen from Fig.\ \ref{fig:KS3} that with pinching 
the sum rule is saturated at smaller $s_{0}$, i.e.\ beginning 
at $s_{0}=2.2$ GeV$^{2}$. In addition the errors from experiment 
are substantially reduced. From Fig.\ \ref{fig:KS3} we would 
extract
\[
f_{\pi}^{2}=0.008\pm0.004\;\mbox{GeV}^{2}\;,
\]
for curve (a), and
\[
f_{\pi}^{2}=0.0084\pm0.0004\;\mbox{GeV}^{2}\;,
\]
for curve (b), to be compared with the experimental value 
$f_{\pi}^{2}|_{EXP}=0.00854\pm0.00005\;\mbox{GeV}^{2}$. 
Curve (a) demonstrates incidentally that it may be dangerous 
to pick up only a small stability region to obtain a prediction 
(here one could choose the region around 2 GeV$^{2}$).


\subsection{The DGLMY sum rule}

\begin{figure}[t]
\begin{center}
\includegraphics[width=0.6\textwidth]{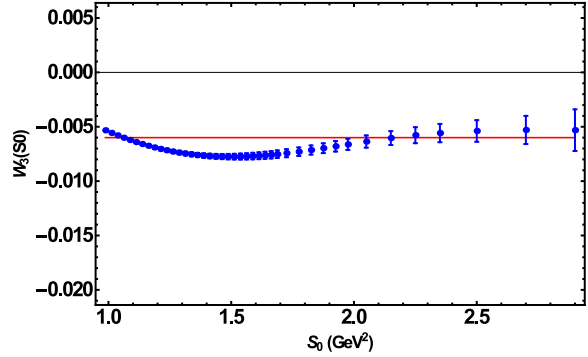}
\caption{\small 
The DGLY sum rule.  
\label{fig:KS4}}
\end{center}
\end{figure}

With the help of the second Weinberg sum rule, the
Das-Guralnik-Low-Mathur-Young (DGLMY)\cite{DGLMY} sum rule can 
also be written in a pinched form, albeit with logarithmic 
pinching,
\[
W_{3}(s) 
\equiv 
\int\limits_{0}^{s_{0}\rightarrow\infty}ds s 
\ln\frac{s}{s_{0}}
\left[\rho_{V}(s)-\rho_{A}(s)\right] 
= -\frac{4\pi f_{\pi}^{2}}{3\alpha}
(m_{\pi^{\pm}}^{2}-m_{\pi^{0}}^{2}) 
\text{ .}
\]
It is seen from Fig.\ \ref{fig:KS4} that the sum rule is 
saturated to the extent expected from the specific kernel. 
The central sum rule result for the mass difference is a few 
percent larger than the experimental value. 


\subsection{The DMO sum rule}

\begin{figure}[h]
\begin{center}
\includegraphics[width=0.6\textwidth]{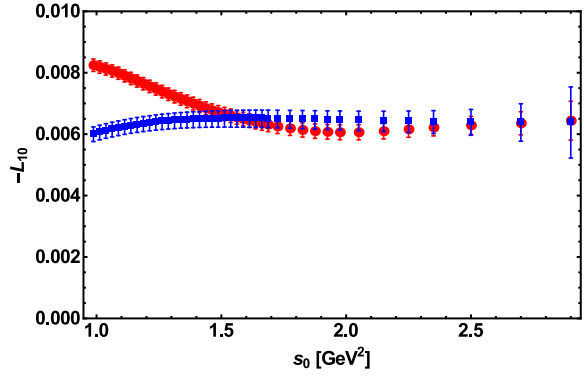}
\caption{\small 
The CHPT constant $-\bar{L}_{10}$ obtained from the pinched 
chiral sum rule for $\bar{\Pi}(0)$ Eq.\ (\ref{M35}). Red dots
correspond to no pinching, blue squares to pinching.  
\label{fig:KS5}}
\end{center}
\end{figure}

For the negative power kernels the FESR are satisfied better 
because the influence of experimental errors is reduced. The 
second Weinberg sum rule is again exploited to generate 
pinching. The corresponding pinched Das-Mathur-Okubo (DMO)\cite{DMO} 
sum rule reads 
\begin{equation}
\bar{\Pi}(0) 
= 4\,\frac{f_{\pi}^{2}}{s_{0}} \, + \, 
\int_{0}^{s_{0}}\,\frac{ds}{s}\, 
\left(1-\frac{s}{s_{0}}\right)^{2} 
\left[\rho_{V}(s)-\rho_{A}(s)\right] 
\ . 
\label{Eq:23} 
\end{equation}
where $\bar{\Pi}(0) = \Pi(0)$ minus the pion pole. In lowest 
order CHPT\cite{DMO} 
\begin{align*}
\bar{\Pi}(0) 
&= 
-8\bar{L}_{10} 
\\
&= 
2\left[\frac{1}{3}f_{\pi}^{2}\langle r_{\pi}^{2} \rangle 
-F_{A}\right] 
\; = 0.052 \pm 0.002 
\;,
\end{align*}
or $\bar{L}_{10} = -(6.33\times10^{-3}\pm0.06)$. Here 
$\langle r^2_\pi \rangle = 0.439 \pm 0.008$ fm$^2$ is the 
electromagnetic radius of the pion\cite{PIONR}, and $F_A = 
0.0119 \pm 0.0001$ the radiative pion decay constant\cite{PDG}. 
Our result is plotted in Fig.\ \ref{fig:KS5}. Numerically we get
\begin{equation}
\bar{L}_{10} 
= 
-(6.5\pm0.1)\times10^{-3}\text{ .} 
\label{M35} 
\end{equation}
It is seen that pinching is hardly necessary for this kernel. 
There is full agreement with alternative 
calculations\cite{Boito1,Pich1} and also with lattice QCD 
calculations\cite{LAT} within their larger uncertainties. 

The FESR for the first derivative of the chiral correlator 
$\Pi^{\prime}(0)$ is related to the $\mathcal{O}$$(p^{6})$ 
counter terms. The pinched sum rule reads 
\begin{equation}
\overline{C}_{87} 
= \Pi^{\prime}(0) + \frac{2f_{\pi}^{2}}{m_{\pi}^{4}} 
= 
\int_{0}^{s_{0}}\left(1-\frac{s^{3}}{s_{0}^{3}}\right) 
\frac{ds}{s^{2}}(\rho_{V}(s)-\rho_{A}(s)) 
\, .
\label{M36} 
\end{equation} 
This value agrees within errors with the value obtained 
in Ref.\cite{Pich1}. We conclude that for negative moments 
the experimental errors at the large energy region of the 
spectral integral become increasingly less important. Pinching
brings no improvement to these FESR. The agreement with 
alternative results is excellent.

\begin{figure}[h]
\begin{center}
\includegraphics[width=0.6\textwidth]{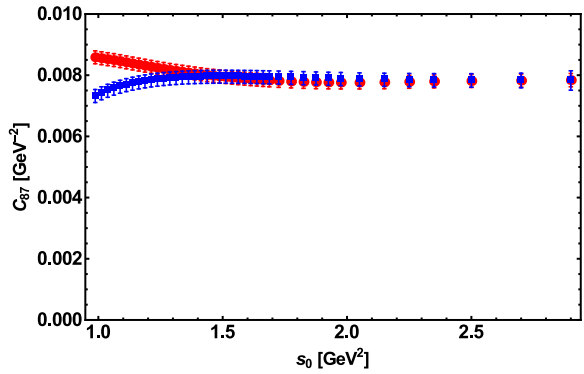}
\caption{\small 
The $O(p^6)$ counterterm $\overline{C}_{87}$ of CHPT according to 
the sum rule Eq.\ (\ref{M36}), squares with pinching, dots 
without pinching. 
\label{fig:KS6}}
\end{center}
\end{figure}


\subsection{Chiral condensates}

Due to the absence of PQCD contributions, it becomes feasible 
to extract chiral condensates with the help of pinched FESR 
to reasonable accuracy. The sum rule for the dimension 6 
chiral condensate reads 
\begin{equation}
\langle{\mathcal{O}}_{6}\rangle 
= 
- 2\,f_{\pi}^{2}\,s_{0}^{2}\, + \, s_{0}^{2}\, 
\int_{0}^{s_{0}}ds\,\left(1-\frac{s}{s_{0}}\right)^{2}\; 
\left[\rho_{V}(s)-\rho_{A}(s)\right] 
\;. 
\label{M43} 
\end{equation} 
\begin{figure}[h]
\begin{center}
\includegraphics[width=0.6\textwidth]{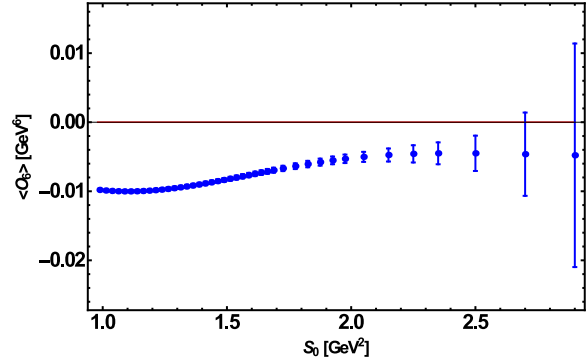}
\caption{\small 
The chiral condensate of dimension $d=6$ from the pinched 
chiral sum rule Eq.\ (\ref{M43}). 
\label{fig:KS7}}
\end{center}
\end{figure}
\begin{figure}[h]
\begin{center}
\includegraphics[width=0.6\textwidth]{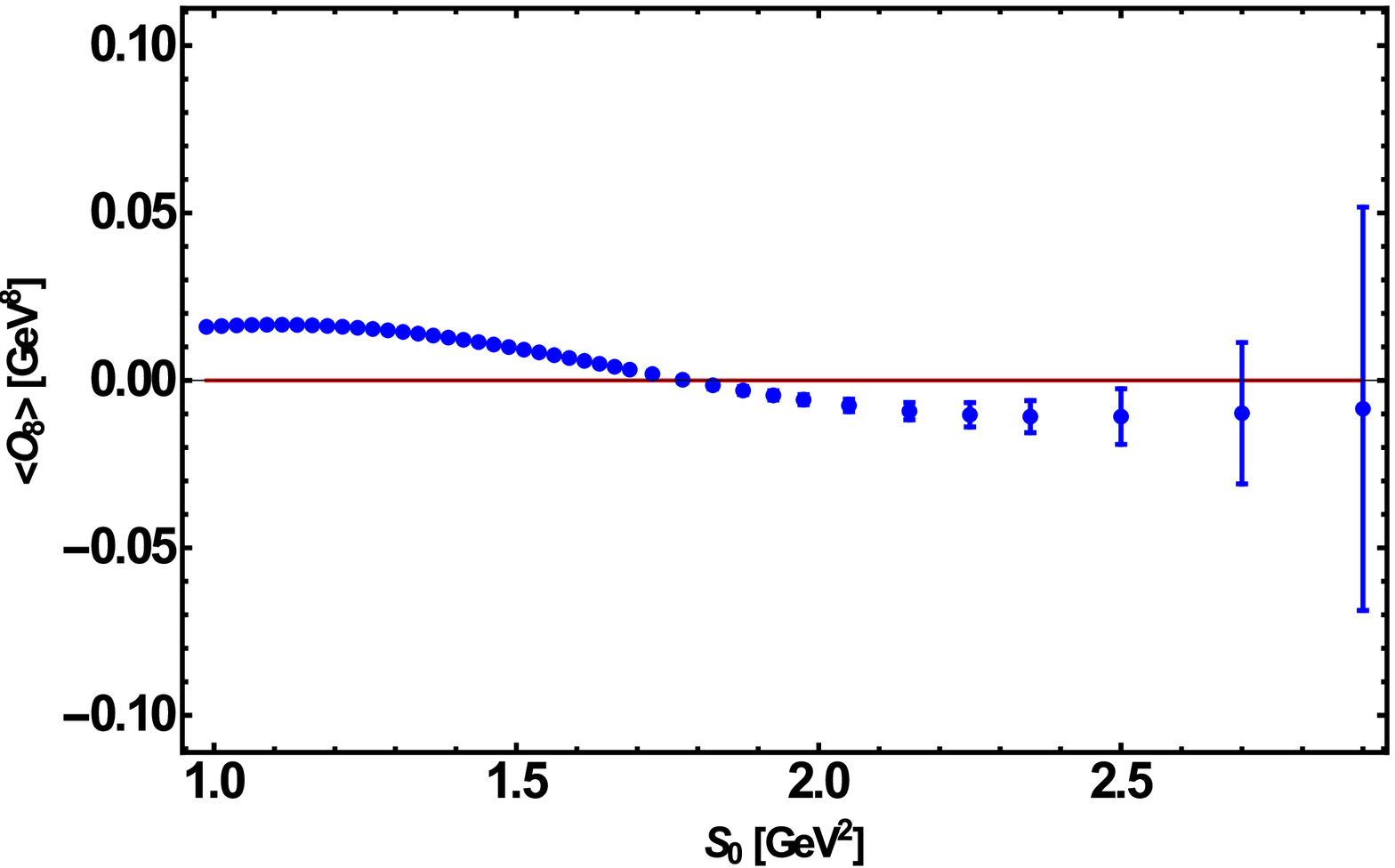}
\caption{\small 
The chiral condensate of dimension $d=8$ from the pinched 
chiral sum rule Eq.\ (\ref{M47}). 
\label{fig:KS8}}
\end{center}
\end{figure}
The results are plotted in Fig.\ \ref{fig:KS7}. From this FESR 
we extract 
\begin{equation}
\langle{\mathcal{O}}_{6}\rangle 
= 
-(5.0\,\pm\,0.7)\times10^{-3}\;{\mbox{GeV}}^{6} 
\, . 
\label{M46} 
\end{equation}
By the same token, we can write a pinched FESR for the 
dimension 8 chiral condensate,
\begin{equation}
\langle{\mathcal{O}}_{8}\rangle 
= 
16f_{\pi}^{2}\,s_{0}^{3} 
- 3\,s_{0}^{4}\,\bar{\Pi}(0) 
+ s_{0}^{3}\, 
\int_{0}^{s_{0}}\frac{ds}{s}\,\left(1-\frac{s}{s_{0}}\right)^{3} 
(s+3\,s_{0})\left[\rho_{V}(s)-\rho_{A}(s)\right]  
\; .
\label{M47} 
\end{equation}
The results are plotted in Fig.\ \ref{fig:KS8}. As expected 
the dimension 8 condensate can be extracted from the sum rule 
only with larger error. We find
\[
\langle{\mathcal{O}}_{8}\rangle 
= 
-(9.0\,\pm\,5.0)\times10^{-3}\;{\mbox{GeV}}^{8} 
\; .
\]


\section{Non-Chiral Sum Rules}

\begin{figure}[h]
\begin{center}
\includegraphics[width=0.6\textwidth]{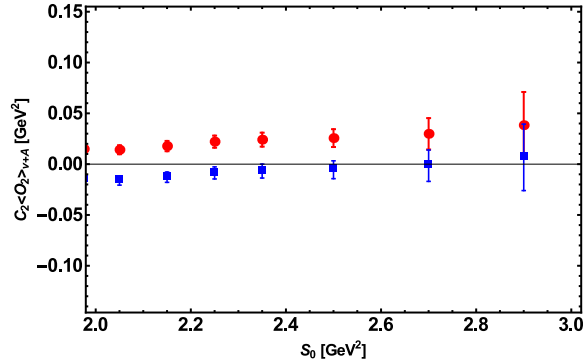}
\caption{\small 
$\langle{\mathcal{O}}_2\rangle$ from the sum rule Eq.\ (\ref{M51}) 
for $\alpha_s=0.354$ (dots)\ and $\alpha_s=0.328$ (squares). 
\label{fig:KS9}}
\end{center}
\end{figure}

When one considers QCD sum rules for the vector and the axial 
vector separately, the perturbative correlators contribute. 
To compare with QCD orthodoxy, we should verify that there is 
no dimension 2 condensate. This can be done with the sum rule
\begin{equation}
\frac{1}{4\pi^{2}}\langle \mathcal{O}_{2} \rangle  
= 
-\int_{s_{thr}}^{s_{0}}ds
\left[
\begin{array}
[c]{c} 
\rho_{V}(s)\\
\rho_{A}(s)
\end{array}
\right] 
- \left[
\begin{array}
[c]{c} 
0\\
1
\end{array}
\right] 
2f_{\pi}^{2}+s_{0}\frac{1}{4\pi^{2}}M_{0}(s_{0}) 
\ . 
\label{M51}
\end{equation}
We evaluate the moments with contour improved QCD\cite{CIPT}. 
As an input we use
\[
\alpha_{s}(m_{\tau}^{2})=0.341\pm0.013
\]
obtained in a recent analysis\cite{Pich2} based on similar 
assumptions as the ones made here. The results for the $V+A$ 
correlator for the two extreme values of $\alpha_{s}(m_{\tau}^{2})$ 
are plotted in Fig.\ \ref{fig:KS9}. It is seen that within the 
experimental errors there is no evidence for the existence of 
a non-vanishing dimension 2 operator. We use this fact to
construct pinched sum rules.

\begin{figure}[h]
\begin{center}
\includegraphics[width=0.6\textwidth]{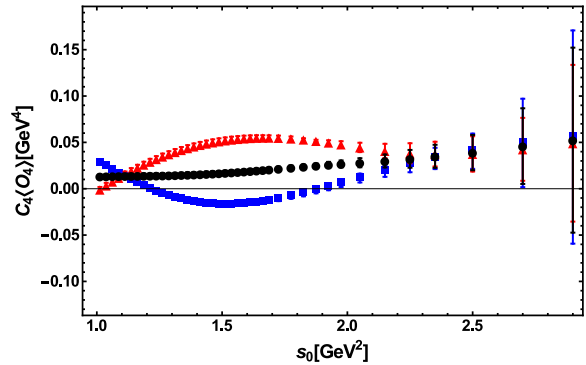}
\caption{\small 
$\langle{\mathcal{O}}_4\rangle$ from the sum rule Eq.\ (\ref{M52}) 
for the $V$ (triangles), $A$ (squares) and $\frac{1}{2}(V+A)$ 
(dots) correlator. 
\label{fig:KS10}}
\end{center}
\end{figure}

As a first application, we consider the dimension 4 condensate 
for the vector and the axial vector channel
\begin{align}
\frac{1}{4\pi^{2}}\langle \mathcal{O}_{4}^{V,A} \rangle  
&= 
s_{0}\int_{s_{thr}}^{s_{0}}ds\left(1-\frac{s}{s_{0}}\right) 
\left[
\begin{array}
[c]{c} 
\rho_{V}(s)\\
\rho_{A}(s)
\end{array}
\right]  -\left[
\begin{array}
[c]{c} 
0\\
1
\end{array}
\right] 
2f_{\pi}^{2} 
\nonumber 
\\
& - s_{0}^{2}\frac{1}{4\pi^{2}}[M_{0}(s_{0})-M_{1}(s_{0})] 
\; 
\label{M52} 
\end{align}
In the chiral limit, i.e.\ for massless quarks, the gluon 
condensates for the $V$ and $A$ correlators should be equal. 
Figure \ref{fig:KS10} shows beautifully how this is
realized by the pinched sum rules. Note that for lower $s_{0}$, 
the duality violations tend to cancel between $V$ and $A$. 
From the figure we extract a value for the gluon condensate 
\[
\langle \mathcal{O}_{4}^{V,A} \rangle 
= \frac{\pi^{2}}{3}\, 
\langle\frac{\alpha_{s}}{\pi}\,G_{\mu\nu}\,G^{\mu\nu}\rangle 
= (0.017\pm0.012)\ GeV^{2} 
\]
which is positive and equal for $V$ and $A$.

\begin{figure}[h]
\begin{center}
\includegraphics[width=0.6\textwidth]{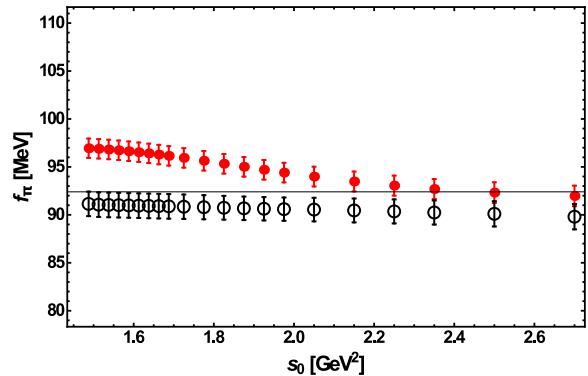}
\caption{\small 
The pion decay constant from the sum rule Eq.\ (\ref{M56}) 
assuming duality (dots) and for a model of DV (circles).
\label{fig:KS11}
}
\end{center}
\end{figure}

\begin{figure}[h]
\begin{center}
\includegraphics[width=0.6\textwidth]{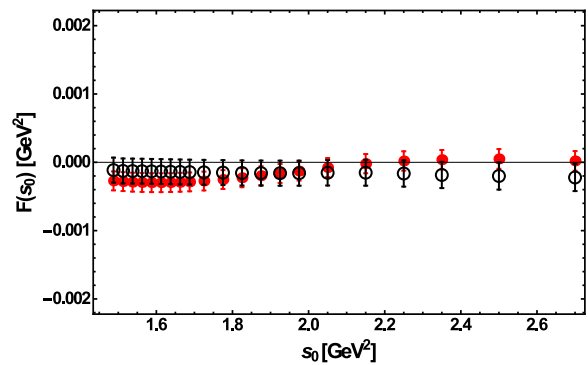}
\caption{\small 
$F(s_0)$ from the sum rule Eq.\ (\ref{M57}) assuming duality 
(squares) and for a model of DV (dots). 
\label{fig:KS12}}
\end{center}
\end{figure}

Given the condensates from independent sources, we can invert 
the sum rule to calculate the pion decay constant 
\begin{align}
2f_{\pi}^{2} 
&= 
-\int_{s_{thr}}^{s_{0}}ds\;\left(1-\frac{s}{s_{0}}\right)
\rho_{A}\left(s\right) 
\nonumber 
\\
& - \frac{1}{2\pi i}\oint_{|s|=s_{0}}ds\, 
\left(1-\frac{s}{s_{0}}\right)
4\pi^{2}\,\Pi_{A}^{QCD}(s) 
+ \frac{1}{4\pi^{2}}
\left[\langle\mathcal{O}_{2}\rangle 
- \langle\mathcal{O}_{4}\rangle\right] 
\ . 
\label{M56} 
\end{align}
For popular values of $\langle\alpha_{s}GG\rangle$ the 
contribution of the condensates is negligible. The result is 
plotted in Fig.\ \ref{fig:KS11}. A more sensitive test is 
produced by the vector correlator, 
\begin{align}
F(s_0)  
= 
-\int_{s_{thr}}^{s_{0}}ds\;\left(1-\frac{s}{s_{0}}\right)
\rho_{V}\left(s\right) 
- \frac{1}{2\pi i}\oint_{|s|=s_{0}}ds\, 
\left(1-\frac{s}{s_{0}}\right)
4\pi^{2}\,\Pi_{V}^{QCD}(s) 
\ . 
\label{M57} 
\end{align}
In this case the left-hand 
side of Eq.\ (\ref{M56}), which we define to be $F(s_{0})$, 
vanishes. We plot the results in Fig.\ \ref{fig:KS11} and 
Fig.\ \ref{fig:KS12} both for our duality approach and for a 
popular model of DV\cite{Boito2} (see Ref.\ \cite{DHSS2} for 
details). As expected the results of the latter are better for 
small $s_{0}$. For the contour radius larger than about $2.2$ 
GeV$^{2}$, however, both approaches are equivalent.



\section{Conclusions}

Duality with pinching works well in $\tau$-decay for 
$s_{0}\gtrsim2.3$ GeV$^{2}$ to $m_{\tau}^{2}$ for all 
observables where the answer is known. The highlights of 
our analysis are the early saturation of the Weinberg sum rules,
the equality of the gluon condensate extracted from the $V$ and 
$A$ spectral functions, the extraction of chiral condensates 
and of some low-energy constants of CHPT. We use only linear 
kernels because of unknown higher dimensional condensates and 
possibly enhanced DV for higher powers. Within the experimental 
errors DV become unobservable for the pinched weights considered.
It is essential that all QCD constraints regarding condensates 
are incorporated. It remains to be seen if sum rules with higher 
powers of the pinch kernel can be satisfied in the straightforward 
FESR approach. The answer to this question would require a 
systematic analysis allowing for higher dimensional vacuum 
condesates and their radiative corrections.


\section*{Acknowledgements}

We thank D.\ Boito, M.\ Golterman, K.\ Maltman and S.\ Peris for 
discussions about their work. One of us (K.S.) thanks K. Maltman 
for a correspondence. We acknowledge financial support by the 
Deutsche Forschungsgemeinschaft, the Mainz Institute for 
Theoretical Physics (MITP), and the National Research
Foundation (South Africa).


\end{document}